\documentclass[aps, 10pt, pra, twocolumn, a4paper, superscriptaddress, longbibliography]{revtex4-1}
\usepackage{amsmath, amssymb, amsthm}
\usepackage[english]{babel}
\usepackage{xcolor}
\usepackage{graphicx}
\usepackage{physics}
\usepackage{tikz}
\usepackage{hyperref}
\usepackage{comment, ulem}
\usepackage[utf8]{inputenc}

\begin{document}

\def\lptms{Universit\'e Paris-Saclay, CNRS, LPTMS, 91405, Orsay, France.}   
\def\enslyon{Univ.~Lyon, Ens de Lyon, CNRS, Laboratoire de Physique, F-69342 Lyon, France.}   
\def\inoTN{INO-CNR BEC Center and Dipartimento di Fisica, Universit\`a di Trento, 38123 Povo, Italy.}
\def\cuny{Physics Department, the City College of New York, NY 10031, USA and The Graduate Center of CUNY, New York, NY 10016, USA}

\title{Measurable fractional spin for quantum Hall quasiparticles on the disk}

\author{Tommaso~Comparin} 
\email{tommaso.comparin@ens-lyon.fr}
\affiliation{\enslyon}
\author{Alvin~Opler}
\affiliation{\lptms}
\author{Elia~Macaluso}
\affiliation{\inoTN}
\author{Alberto~Biella}
\affiliation{\lptms}
\affiliation{\inoTN}
\author{Alexios~P.~Polychronakos}
\affiliation{\cuny}
\author{Leonardo~Mazza}
\email{leonardo.mazza@universite-paris-saclay.fr}
\affiliation{\lptms}

\date{\today}

\begin{abstract} 
We study the spin of the localised quasiparticle excitations of lowest-Landau-level quantum Hall states defined on a disk.
The spin that we propose satisfies the spin-statistics relation and can be used to reconstruct the topological geometric phase associated to the exchange of two arbitrarily chosen quasiparticles.
Since it is related to the quadrupole moment of the quasiparticle charge distribution, it can be measured in an experiment and could reveal anyonic properties in a way that is complementary to the interferometric schemes employed so far.
We first discuss our definition for the quasiholes of the Laughlin state, for which we present a numerical and analytical study of our spin, and
we proceed with a discussion of several kinds of quasiholes of the Halperin 221 state.
Finally, we discuss the link between our spin and the adiabatic rotation of the quasiparticles around their axis and demonstrate that our spin obeys the spin-statistics relation.
\end{abstract}

\maketitle

\section{Introduction}

In recent years, an increasing amount of attention has been devoted to the study
of \textit{anyons}, quasiparticles which display a quantum statistics that is
neither bosonic nor fermionic~\cite{Leinaas_1977, Goldin_1981, Wilczek_Book, Lerda_Book, Rao_AnyonPrimer, Preskill_2004, Khare_Book, Ouvry_2009}. 
On top of the tremendous scientific interest for these objects, it has also been proposed that they could play a key role in novel quantum computing schemes, dubbed topological, and thus trigger the development of innovative technologies~\cite{Kitaev_2003, Nayak_2008}.

The simplest model for an anyon is that of a charge-flux composite, made of a charge $q$ orbiting around an infinitesimally-thin solenoid with magnetic flux $\Phi$~\cite{Wilczek_1982_A, Wilczek_1982_B}.  
Historically, this represented a first concrete instance of a quantum object living in two spatial dimensions characterised by fractional properties. 
One key result of the studies on this model is that the spin $S$ of the charge-flux composite is related to the geometric topological phase  $e^{i \theta}$ picked up by the system wavefunction once two composites are exchanged in a counter-clockwise fashion. The link between the two concepts is expressed by $e^{2 \pi iS} = e^{i \theta}$, or, equivalently, by $S= \theta/2 \pi+ \mathbb Z$, and constitutes an instance of the \textit{spin-statistics theorem}~\cite{Polychronakos_1987, Balachandran_1990}.

The first experimental setup with quasiparticle excitations to be described in terms of anyons is the two-dimensional electron gas in presence of a strong and perpendicular magnetic field~\cite{Halperin_1984, Thouless_1985, goerbig2009quantum, tong2016lectures}.
Starting from the pioneering work by Laughlin on the fractional quantum Hall effect (FQHE), who introduced several model wavefunctions describing the quasihole and quasielectron excitations at certain filling factors~\cite{Laughlin_1983}, an uncountable series of works have discussed the fractional charge and statistics of these objects, which are now well assessed~\cite{Arovas_1984, Feldman_2021}.

However, as we have seen, the notion of anyon is deeply linked to that of fractional spin. 
The standard definition of spin of an anyon is in terms of the statistical phase acquired by the system when an anyon is exchanged with its anti-anyon, which is a quasiparticle with which it can annihilate and fuse to the vacuum~\cite{Preskill_2004}.
In the context of the FQHE, the definition of a quasiparticle spin that satisfies the spin-statistics theorem has turned out to be problematic when the wavefunction is defined on a disk, and it has been studied only when the problem is defined on a spherical surface or on a cylinder.
Concerning the sphere, a few pioneering works have shown that when the wavefunction of the FQHE is defined on a spherical surface and a quasiparticle is adiabatically moved along a closed path, an anomalous rotation appears that can be interpreted as a fractional spin and that is due to the coupling of the spin to the curvature of the sphere~\cite{Li_1992, Wen_1992, Li_1993, Lee_1994, Einarsson_1995, read2008quasiparticle,Gromov_2016}. 
In fact, even in classical systems it is well known that the parallel transport of a vector along a generic path on the spherical surface generates a rotation. 
For what concerns the cylinder, instead, the recent technique of momentum polarisation has provided a tool to access the quasiparticle spin via an analysis of the entanglement properties of the quantum state~\cite{Zaletel_2013, Tu_2013, Zaletel_2014}.

In this article we propose and study a fractional spin for the quasiparticle excitations of an incompressible FQHE state. We will focus on the case of the lowest Landau level (LLL), but generalisations are straightforward. 
The definition is well-suited for states that are defined on a planar surface, such as a disk, which was so far an open problem; but it could also be used on a cylinder, where it would complement the aforementioned momentum-polarisation technique.
We test it with the quasiholes of the Laughlin state (at filling $1/m$, with $m=2$, $3$ and $4$)~\cite{Laughlin_1983} and of the Halperin $221$ state~\cite{Halperin_1983}, where the numerical computation for large systems is made possible by the use of the plasma analogy~\cite{Laughlin_1983, Halperin_1983, goerbig2009quantum, tong2016lectures}.
In the Laughlin case, we also find an analytical proof, and the result agrees with that derived in Refs.~\cite{Li_1992, Lee_1994, Einarsson_1995, Gromov_2016} on the sphere.

The interest of our fractional spin is twofold.
First of all, it satisfies a generalised spin-statistics relation.
This result is well confirmed by our analytical argument and numerical simulations in all situations that we studied: the values that we obtain for the fractional spin of one and of two quasiholes can be combined according to standard recipes to obtain the exchange phase of the quasiholes. 
Secondly, our definition has an experimental relevance because it is only a function of the density profile of the liquid in the vicinity of the quasiparticle. Given that the experimental detection of fractional statistics is a delicate subject, our work could foster experimental studies that are complementary to the existing ones, as it is particularly suitable for measurements performed with ultra-cold gases or quantum fluids of light~\cite{Tai_2017, Aidelsburger_2018, Cooper_2019, Chalopin_2020, Jamadi_2020, Clark_2020}. As an additional advantage, our scheme does not require the study of both the quasiparticle and its anti-quasiparticle, as required by the general definition: the study of clusters of quasiparticles of the same kind is sufficient to predict the desired statistical phases.

The definition of a quasiparticle spin for a FQHE state defined on a planar surface has long been debated, and several works have discussed this possibility.
The quantity that we are proposing is associated to the quasiparticle angular momentum, and was already mentioned in Ref.~\cite{Sondhi_1992} (the statement is reproposed in Ref.~\cite{Einarsson_1995}), but the authors claimed that it should be equal to zero because of the constant-screening sum rule. Our explicit calculation shows that this is not the case. Although the calculation is very demanding from a numerical viewpoint (it is associated to the quadrupole moment of the anyonic charge density), results are not compatible with zero; within error bars, they are compatible with fractional values that are consistent with our analytical predictions. 
We mention also a series of articles that have studied the properties of the second moment of the depletion density of fractional quasiparticles, which is shown to be related to the conformal dimension; the spin-statistics relation has not been discussed explicitly~\cite{Can_2015, Can_2016, Schine_2019}.

Nonetheless, we could perform our calculations only in some specific cases.
For this reason, in the second part of the article we present a general argument supporting our findings;
the basic idea is that a quasiparticle that is at rest in a rotating reference frame performs a motion in the laboratory frame that includes a rotation around itself.
We identify the geometric phase picked up in this process and link it to the fractional spin of the quasiparticle; when discussing two quasiparticles, we recover the aforementioned fractional spin-statistics relation in full generality.  This derivation also provides additional insight into previous works where the braiding phase for quasiparticle excitations was linked to their depletion profiles~\cite{Umucalilar_2018, Macaluso_2019, Macaluso_2020}.
Whereas in those previous works the emphasis was on the statistics of quasiparticles, here we perform a step further and link it to their fractional spins, which was not discussed in those works.

This article is organised as follows. 
In Sec.~\ref{Sec:GeneralisedSpinStatistics} we present our definition of quasiparticle spin and we compute it for the quasiholes of the Laughlin state, for which we show that it satisfies the generalised spin-statistics relation. Our results are benchmarked against the existing literature on the fractional spin of Laughlin quasiholes, and they agree with those in Refs.~\cite{Li_1992, Einarsson_1995}.
In Sec.~\ref{Sec:Halperin221} we study the Halperin 221 state and compute the fractional spin of its quasiholes using our definition. We use these values to infer the statistical phase of the quasiholes, and benchmark our results with the data presented in Ref.~\cite{Jaworowski_2019}; again, our findings are perfectly compatible with a spin-statistics relation.
Further comments on the link with the standard definition of topological spin are presented in Sec.~\ref{Sec:AFI}, where we discuss another model of quasiparticles.
In the second part of the article (Sec.~\ref{Sec:AdiabaticRotation}), we employ a more abstract approach to discuss the nature of our fractional spin.
We present a scheme to rotate the quasiholes and to compute the geometric phase picked up during the process. Once this approach is applied to a single quasihole, it leads in a very natural  way to our definition of fractional spin (see Sec.~\ref{Sec:OneQH}). 
When we specify our calculation to two quasiholes, we obtain information on the statistics of the quasiholes and establish a spin-statistics relation (see Sec.~\ref{Sec:TwoQH}).
Our conclusions and a final discussion on the significance of our results are in Sec.~\ref{Sec:Conclusions}.

\section{Quasiparticle spin}
\label{Sec:GeneralisedSpinStatistics}

In this section we present the definition for the spin of a quasiparticle excitation of a many-body state defined on a disk in the LLL. We compute it for the quasiholes of the Laughlin state and show that although the spin that we propose does not agree with other earlier definitions, (i) it satisfies a correct spin-statistics relation and (ii) it is a measurable quantity.

\subsection{Definition}

We consider a two-dimensional system composed of $N$ quantum particles (they can be bosons or fermions) in the presence of a perpendicular magnetic field $B$; for all particles we use the complex coordinates $z=x+iy$.
We assume that the system realizes an incompressible FQHE state in the LLL.
We also assume that an external pinning potential localises a quasiparticle at position $\eta$, whose distance from the boundary is much larger than any relevant length-scale of the problem. 
The density profile is $\rho_\eta(z)$, and if we neglect boundary effects and assume rotational invariance of the quasiparticle it is just a function of $|z-\eta|$.

We propose the following definition of the quasiparticle spin,
\begin{equation}
 J =  \int_{\mathcal A} \left( \frac{|z-\eta|^2}{2 \ell_B^2} -1 \right)  \left( \rho_\eta(z) - \bar{\rho} \right) d^2z,
 \label{Eq:Def:Spin}
\end{equation}
where $\ell_B = \sqrt{\hbar/(eB)}$ is the magnetic length, $e>0$ is the elementary charge, $\bar{\rho}$ is the bulk density (far from the boundaries, and in the absence of quasiparticles), and we set $\hbar=1$.
The integral is extended over a large region $\mathcal A$ around the quasiparticle, which should be chosen in such a way to avoid the boundary. 
In practice, when performing a numerical simulation, we will consider a disk geometry, place the quasihole at the centre of the disk ($\eta=0$) and integrate over a circle of radius $R$ smaller than the sample radius.
The spin $J$ in Eq.~\eqref{Eq:Def:Spin} is a measure of the excess of angular momentum associated to the presence of the quasiparticle with respect to the homogeneous situation, and it follows from the definition of the canonical angular-momentum operator in the LLL (see Appendix~\ref{App:Angmom}). Alternatively, we can say that the spin of the quasiparticle is a subleading contribution to the angular momentum of the state that includes the quasiparticle. 
We refer to Refs.~\cite{Can_2015, Can_2016, Schine_2019} for previous studies of the second moment of the FQHE quasiparticles.

Moreover, $J$ is a measurable quantity: the definition shows that it is a function of the net total charge and of the quadrupole moment of the depletion density of the quasiparticle,
\begin{equation}
 d_{\eta}(z) = \rho_\eta(z) - \bar{\rho}.
\end{equation}
As we have mentioned in the introduction, this can in principle be measured with an experimental platform able to reconstruct the density profile of the quantum system in the vicinity of $\eta$.

\subsection{The quasiholes of the Laughlin state}

The spin of a quasiparticle has already been computed for several FQHE states defined on a sphere. 
Our definition is different, and as a first consistency check it is important to compare it with the existing results, mainly related to the Laughlin state. 
We consider the filling factor $\frac 1m$ and the wavefunction with quasiparticle composed of $q$ quasiholes centered in $\eta$:
\begin{equation}
 \Psi_{m, q}(z_i, \eta) \propto \prod_i (z_i-\eta)^q  \times \prod_{i<j}(z_i-z_j)^m e^{- \frac{\sum_i |z_i|^2}{4 \ell_B^2}}.
\label{Eq:Laughlin}
\end{equation}
The spin of this quasiparticle presented in Ref.~\cite{read2008quasiparticle} is:
\begin{equation}
 s_{m,q} = - \frac{q^2}{2m};
 \label{Eq:Spin:Laughlin}
\end{equation}
This value is the one associated to the standard formulation of the spin-statistics theorem, as we are going to discuss here below.

The authors of Refs.~\cite{Li_1992, Lee_1994, Einarsson_1995} compute the spin of the $q=1$ quasihole (``qh'') and quasielectron (``qe'', wavefunction not shown here) by moving adiabatically a quasiparticle on the surface, and obtain:
\begin{equation}
 J^*_{m,qh} = - \frac{1}{2m}+ \frac 12;
 \quad 
 J^*_{m,qe} = - \frac{1}{2m}- \frac 12.
 \label{Eq:Spin:Einarsson}
\end{equation}
The discussion in Ref.~\cite{Einarsson_1995} explain in an enlightening way that these results are consistent with $s_{m,1}$ in Eq.~\eqref{Eq:Spin:Laughlin} because $s_{m,1} = \frac 12 \left( J^*_{m,qh}+J^*_{m,qe}\right)$, which is the only quantity that is relevant for establishing a spin-statistics relation. 
We will address these problems in the next subsection~\ref{SubSec:Spin:Statistics}; we devote the rest of this subsection to the calculation of our spin.

\begin{figure*}[t]
\includegraphics[width=\textwidth]{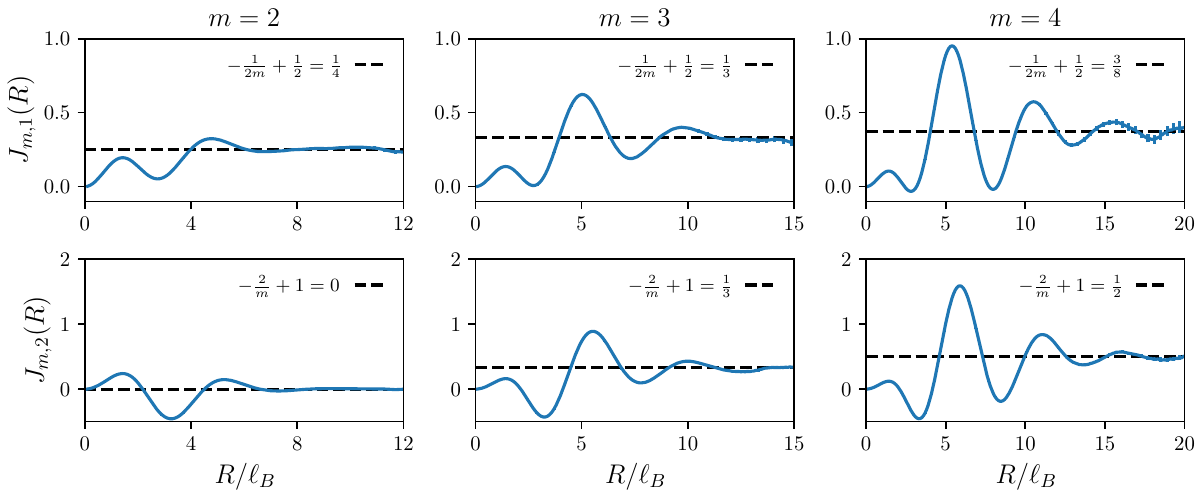}
\caption{
\textbf{Fractional spin of $q$ quasiholes of a Laughlin state.}
Panels in first (second) row show $J_{m,q}(R)$ for $q=1$ ($q=2$), computed via Monte Carlo sampling and Eq.~\eqref{Eq:Spin_with_cutoff}  (blue solid lines), for Laughlin states with filling $1/m$  (see column titles) and $N=200$. Horizontal dashed lines	represent the saturation values $J_{m,q}$ reported in Tab.~\ref{Table:Laughlin}. 
Error bars are statistical uncertainties, and data are only shown for cutoff radii $R$ where boundary effects are negligible and Monte Carlo sampling is reliable.}
\label{Fig:Laughlin}
\end{figure*}

\begin{table}[t]
 \begin{tabular}{c || c | c | c || c | c || c | c}
  \phantom{i}$m$\phantom{i} & \phantom{i}$J_{m,1} $\phantom{i} &\phantom{i}$s_{m,1} $\phantom{i} & \phantom{i}$J^*_{m,qh} $\phantom{i} & \phantom{i}$J_{m,2} $\phantom{i} & \phantom{i}$s_{m,2} $\phantom{i} & \phantom{i}$e^{i\Phi_{ex}}$\phantom{i} & \phantom{i}$e^{i\varphi_{ex}}$\\
  \hline \hline 
  2 & $\frac 14$ & $- \frac 14$ & $\frac 14$ & $0$ & $-1$ &
  $e^{ i \frac{\pi}2}$ & $e^{ i \frac{\pi}2}$\\
  3 & $\frac 13$ & $-\frac 16$ & $ \frac 13$ &$ \frac 13$ & $- \frac 23$ & $e^{i \frac\pi3}$ & $e^{i \frac\pi3}$\\
  4 & $\frac 38 $ & $- \frac 18$ & $ \frac 38$ & $ \frac 12$ & $-\frac 12$ & $e^{ i \frac{\pi}4 }$ & $e^{ i \frac{\pi}4 }$
 \end{tabular}
 \caption{\textbf{Fractional spin and exchange phase of the quasiholes of a Laughlin state} computed with different techniques. For the calculation of $J_{m,q}$ according to the definition in Eq.~\eqref{Eq:Def:Spin}, we give the rational number compatible with the simulation (horizontal dashed line in Fig.~\ref{Fig:Laughlin}), while $s_{m,q}$ and $J^*_{m,qh}$ are defined in Eqs.~\eqref{Eq:Spin:Laughlin} and~\eqref{Eq:Spin:Einarsson} and taken respectively from Refs.~\cite{read2008quasiparticle} and~\cite{Einarsson_1995}.}
 \label{Table:Laughlin}
\end{table}

In order to compute the spin of the Laughlin quasihole according to our definition and to compare it to Eqs.~\eqref{Eq:Spin:Laughlin} and~\eqref{Eq:Spin:Einarsson}, we reconstruct the density profile of the state $\Psi_{m,q}$ using the plasma analogy. We consider the values $m = 2,3,4$, and we perform a Monte Carlo sampling of $|\Psi_{m,q}|^2$ using the Metropolis algorithm with either one quasihole ($q=1$) or two quasiholes ($q=2$) placed at the centre of the disk ($\eta=0$). 
We then reconstruct the density profile $\rho_{0}(z)$ and the associated depletion density $d_{0}(z)=\rho_{0}(z)- 1/(2\pi m \ell_B^2)$, which only depends on the distance $r=|z|$ from the centre of the disk.
We study the integral
\begin{equation}
J_{m,q}(R) =
2 \pi \int_0^R \left(\frac{r^2}{2 \ell^2_B} -1\right) d_{0}(r) r dr,
\label{Eq:Spin_with_cutoff}
\end{equation}
up to a cut-off radius $R$; in the absence of the boundary, we expect this quantity to converge to the spin of the particle $J_{m,q}$ in Eq.~\eqref{Eq:Def:Spin}, for large $R$.

Our results are shown in Fig.~\ref{Fig:Laughlin}. The calculation of the second moment of $d_0(r)$ is particularly demanding, and even with our high-precision numerical data (obtained with up to $\approx 10^{10}$ global Monte Carlo moves) the estimate of $J_{m,q}(R)$ becomes unreliable at large $R$. Nevertheless, we clearly identify a saturation value $J_{m,q}$ at $R/\ell_B \approx 10-20$, reached after a few damped oscillations. The fractional spin $J_{m,q}$ that we extract are compatible with simple rational numbers, as summarised in Table~\ref{Table:Laughlin}, where we also compare them with Eqs.~\eqref{Eq:Spin:Laughlin} and~\eqref{Eq:Spin:Einarsson}. 

It is easy to see that $s_{m,q}$ and $J_{m,q}$ are different;
if we consider $q=1$, we observe that $J_{m,1}-s_{m,1} = 1/2$: our spin has a systematic shift with respect to Eq.~\eqref{Eq:Spin:Laughlin}. 
On the other hand, $J_{m,1} = J^*_{m,qh}$; in our case, however, with the numerical tools at our disposal we cannot compute the spin of Laughlin quasielectron and check the consistency relation proposed in Ref.~\cite{Einarsson_1995}.

Moving to the case $q=2$, we observe that $J_{m,2} = 4 J_{m,1} \mod 1$. This is an interesting relation that parallels the analogous cluster property satisfied by the spin defined in Eq.~\eqref{Eq:Spin:Laughlin}, namely the scaling as $q^2$~\cite{Thouless_1985}. In fact, the general theory of Abelian anyons predicts that when grouping together $n$ anyons, one obtains an anyon whose spin has been multiplied by a factor $n^2$.
The fact that in our case the equality is satisfied modulus $1$ can be accepted once we observe that in order to establish a generalised spin-statistics relation (our ultimate goal) we are only interested in $2 \pi$ rotations, so that unity terms are inessential. 

\subsubsection{An analytical calculation of $J_{m,q}$}\label{SubSubSec:AnalyticalProof}
\label{Laughlin:Proof}

Starting from a few simple considerations,
it is possible to extend our numerical results and to
show in an analytical way that 
\begin{equation}
J_{m,q} = - \frac{q^2}{2m}+ \frac{q}{2} .
\end{equation}

Let us consider the radial density profile of the Laughlin state, $\rho(r)$, which is flat in the bulk, with value $\bar \rho = \frac{1}{ 2\pi m \ell_B^2} $, and decreases to zero in the vicinity of the classical radius of the droplet, $r_0 = \sqrt{N/(\pi \bar \rho)}$, defined by $\int_0^{r_0} \bar \rho \, 2 \pi r dr=N$. 
In the large-$N$ limit, the profile of the gas is universal (see Appendix~\ref{App:Boundaries}), apart from negligible corrections, and it is thus possible to express it as:
\begin{equation}
 \rho(r) = \rho_{\bar \rho, r_0}(r) 
\end{equation}
where the universal function $\rho_{\bar \rho, r_0}$ \textit{only} depends on the bulk value $\bar \rho$ and on the classical radius $r_0$. 

We know that $\int_0^{\infty}\rho(r) 2 \pi r dr = N$ and that $\int_0^{\infty}\frac{r^2}{2 \ell_B^2}\rho(r) 2 \pi r dr = N+L$, where $L$ is the angular momentum of the state (see for instance Appendix~\ref{App:Angmom}); for a Laughlin state $L=m \frac{N(N-1)}{2}$. We can conclude that
\begin{subequations}
 \begin{align}
  \int_0^{\infty} \rho_{\bar \rho, r_0}(r) 2 \pi r dr =& N \\
  \int_0^{\infty} \frac{r^2}{2} \rho_{\bar \rho, r_0}(r) 2 \pi r dr =& N + m \frac{N(N-1)}{2} .
 \end{align}
\end{subequations}
In fact, we can say more than this. The profile $\rho_{\bar \rho, r_0}(r)$ is only a function of $\bar \rho$ and $r_0$. As such,  we expect the integral to depend only on the two parameters $\bar \rho$ and $r_0$. 
Using the explicit expressions for these parameters given above,
we can thus conclude the two following important properties of the universal part of the density profile of a Laughlin state:
\begin{subequations}
 \begin{align}
  \int_0^{\infty} \rho_{\bar \rho, r_0}(r) 2 \pi r dr =& \pi  \, \bar \rho \, r_0^2; \\
  \int_0^{\infty} \frac{r^2}{2} \rho_{\bar \rho, r_0}(r) 2 \pi r dr =& \pi  \, \bar \rho \, r_0^2 + \frac 12 \frac{r_0^2}{2 \ell_B^2} \left( \pi  \, \bar \rho \, r_0^2 -1 \right).
 \end{align}
\end{subequations}


Equipped with these results, we can now characterise the depletion density of the quasihole. We assume that when we add $q$ quasiholes in the center, the density profile of the system becomes:
\begin{equation}
 \rho(r) = d_0(r) + \rho_{\bar \rho, r_q}(r). 
\end{equation}
The depletion density $d_0(r)$ has a charge $-q/m$, and it describes the deviations of the density profile from the homogeneous case; note that $d_0(r) \sim e^{- r/ r_0}$ for large $r$. Concerning the boundary, we just assume that it maintain the same universal form of the homogeneous gas, but that it is slightly shifted: the new classical radius $r_q$ is chosen to accommodate the particles that have been pushed away from the bulk and it
is defined by the conservation of charge:
\begin{equation}
 \pi \bar \rho \, r_q^2 = N + \frac{q}{m}.
\end{equation}
Concerning the second moment:
\begin{align}
 \int_0^{\infty} \frac{r^2}{2} \rho_{\bar \rho, r_q}(r) 2 \pi r dr =& \nonumber \\= N + \frac qm + \frac 12 & \left( Nm +  q \right)   \left( N + \frac qm -1 \right).
 \label{Eq:Second:Moment:Depletion:Analytics}
\end{align}
The relation $\int \frac{r^2}{2 \ell_B^2} \rho(r) 2 \pi r dr = N+L$ is still valid, and for a state with $q$ quasiholes the angular momentum is $L = m \frac{N(N-1)}{2}+ qN$,
so that the integral in Eq.~\eqref{Eq:Second:Moment:Depletion:Analytics} should be equal to
\begin{equation}
 N+ m \frac{N(N-1)}{2}+ qN - 
 \int_0^{\infty}\frac{r^2}{2 \ell_B^2}d_0(r) 2 \pi r dr.
\end{equation}
With little algebra, we obtain:
\begin{equation}
J_{m,q} =
 \int \left(\frac{r^2}{2 \ell_B^2}-1 \right) d(r) 2 \pi r dr = - \frac{q^2}{2m}+ \frac{q}{2}
 \label{Eq:Laughlin:J:Proof}
\end{equation}
The values extracted from our numerical simulations are compatible with this general result that only assumes the existence of a universal density profile, the exponential localisation of the quasihole, and the rigid shift of the boundary of the gas once the quasiholes are inserted.

In an early work, Sondhi and Kivelson have proposed our definition for the quasiparticle spin but dismissed it on the basis of the constant-screening sum rule, according to which they computed its value to be zero~\cite{Sondhi_1992}; similar statements are presented in Ref.~\cite{Einarsson_1995}. The present explicit calculation shows a value that is different from zero; a better understanding of the reasons of the disagreement is planned for a future work.

\subsection{Spin-statistics relation}\label{SubSec:Spin:Statistics}

Notwithstanding the numerical differences with $s_{m,q}$, our expression $J_{m,q}$ satisfies the correct generalised spin-statistics relations.
Let us first define what we are speaking about. 
We consider two quasiparticles of kind $a$ and $b$, that fuse together in a quasiparticle of type $c$; each of them is characterised by a fractional spin $J_{\alpha}$, with $\alpha = a$, $b$, $c$. We define the topological \textit{braiding phase} $e^{i { \Phi_{br}} }$ that is acquired by the system when the particle $a$ { encircles} in a counter-clockwise way the quasiparticle $b$ and they are in the fusion channel $c$: the spin-statistics relation states that
\begin{equation}
 e^{i  \Phi_{br}} = e^{-  { 2 \pi} i \left( J_c - J_a - J_b\right)}.
 \label{Eq:Spin:Statistics}
\end{equation}
If the particles are of the same type, i.e.~$a=b$, we can define the topological \textit{exchange phase} that is picked up when the two particles are exchanged in a counterclockwise way:
\begin{equation}
{ e^{i  \Phi_{ex}} = e^{- i \pi \left( J_c - { 2}J_a \right)}.}
 \label{Eq:Spin:Statistics:2}
\end{equation}
The braiding phase and the exchange phase are related by a factor $2$ because an encircling process is topologically equivalent to two exchanges, $\Phi_{ex} = \Phi_{br}/2$.
The standard spin statistics relation is typically obtained by assuming the cluster relation, $J_c = 4 J_a$ so that the statistical parameter $\theta = - \Phi_{ex}$ satisfies $J_a = \theta / (2 \pi) + \mathbb Z$~\cite{Hansson_1992}.

The spin values that are reported in Table~\ref{Table:Laughlin} are sufficient to compute the exchange phase for two identical quasiholes $q=1$ using Eq.~\eqref{Eq:Spin:Statistics:2}.  Depending on whether we use our spin definition or the one in Eq.~\eqref{Eq:Spin:Laughlin}, the exchange phases $e^{i \Phi_{ex}}$ and $e^{i \varphi_{ex}}$ respectively, read:
\begin{equation}
 e^{i\Phi_{ex}} = e^{- i \pi (J_{m,2}-2 J_{m,1})}
 , \quad
 e^{i\varphi_{ex}} = e^{- i \pi (s_{m,2}-2 s_{m,1})}.
\end{equation}
Remarkably, as summarised in Table~\ref{Table:Laughlin}, we obtain $e^{i \Phi_{ex}} = e^{i \varphi_{ex}}= e^{i \frac{\pi}{m}}$: this the known value computed with the explicit and independent calculations of the Berry phase via the plasma analogy~\cite{Arovas_1984}.

By explicit inspection, the spin $J_{q,m}$ is generally compatible with the spin-statistics relation~\eqref{Eq:Spin:Statistics:2} even if it does not satisfy the cluster property because the added term is proportional to $q$.
It is interesting to observe that the prefactor $1/2$ grants an interesting property, namely that $J_{m,2q} = 4 J_{m,q} \mod 1$, so that we can speak of a generalised cluster property.

In summary, when we apply the definition of quasiparticle spin proposed in Eq.~\eqref{Eq:Def:Spin} to the quasiholes of the Laughlin state, we do not reproduce the values obtained in Ref.~\cite{read2008quasiparticle}; our values, obtained with analytical and numerical means, match exactly with the Berry-phase calculations performed on a sphere in Refs.~\cite{Li_1992, Einarsson_1995}. 
We have also shown that the spin $J_{m,q}$ satisfies a generalised cluster property (since it scales as $q^2$) and the generalised spin-statistics relation.

\subsection{Gauge invariance}

Looking at the spin defined in Eq.~\eqref{Eq:Def:Spin}, a natural question is whether the observable that we are proposing is a gauge-invariant quantity. 
So far, all our derivations have been carried out in the symmetric gauge exploiting rotational symmetry; moreover, we have shown that calculating the spin $J$ amounts to computing the excess of angular momentum with respect to the ground state. 
Whereas this has the physical meaning of linking the quasiparticle spin to the generator of rotations, it has the problem that the canonical angular momentum is not a gauge invariant quantity; the generality of our result might thus be questionable.
In this paragraph we show (i) that it is possible to define a gauge-invariant generator of rotations, and (ii) that in our specific case it coincides with the canonical momentum.

We consider a particle with charge $q$, and we consider the position and gauge invariant mechanical momentum operators, $x$, $y$, $\pi_x$ and $\pi_y$, which satisfy the well-known commutation relations (we list here only those that are not zero, and we restore $\hbar$ for clarity): 
\begin{equation}
 [ x, \pi_x] = i \hbar,
 \quad
 [ y,  \pi_y] = i \hbar,
 \quad
 [ \pi_x,  \pi_y] =  i q \hbar B( x,y).
\label{defJ}\end{equation}
The current discussion is general and includes the case of an inhomogeneous magnetic field.
We introduce the gauge-invariant generator of rotations $J$ and we require that it satisfies:
\begin{subequations}\begin{align}
 [ J,  x] = i \hbar  y,
 \qquad &
 [J, y] = - i \hbar x; \\
 [J, \pi_x] = i \hbar \pi_y,
 \qquad &
 [ J,  \pi_y] = - i \hbar \pi_x.
 \label{Eq:Momentum:Rotation:GaugeInv}
\end{align}\end{subequations}
A first possibility would be to use the mechanical angular momentum, $ x \pi_y -  y   \pi_x$, which is gauge invariant; however, a simple check shows that it does not satisfy the Eqs.~\eqref{Eq:Momentum:Rotation:GaugeInv}. We thus set $ J = x \pi_y - y  \pi_x + R$ and try to derive the appropriate gauge-invariant operator $R$. Application of~\eqref{defJ} yields $[ x,  R] = [y, R] = 0$ and
\begin{equation}
 [\pi_x,  R] = i q \hbar B( x, y)  x, \qquad 
 [ \pi_y, R] = i q\hbar B( x, y) y.
\end{equation}
The vanishing commutators with $ x$ and $y$ imply that $R$ is a function of $x$ and $y$ only.
The last two commutators imply:
\begin{equation}
 \frac{\partial R}{\partial x}= -q B(x,y) x,
 \qquad 
 \frac{\partial R}{\partial y}= -q B(x,y) y.
 \label{Eq:Def:R}
\end{equation}
Two main conclusions can now be drawn. The first follows from the consistency relation
$\partial_x \partial_y R = \partial_y \partial_x R$. We obtain:
\begin{equation}
  x \partial_y B  - y\partial_x B = 0.
\end{equation}
Thus, a gauge invariant generator of rotations can be defined only when the magnetic field is rotationally invariant.
The second conclusion follows from the integration of Eqs.~\eqref{Eq:Def:R}: 
\begin{equation}
R(x,y) = -q \Phi (r) +C ~~~\text{with}~~~ \Phi (r) = \int_0^r B(r) r dr .
\end{equation}
That is, $\Phi (r)$ is the magnetic flux inside a disk of radius $r$ measured in units of flux quanta $2\pi$. $C$ is an arbitrary
constant, which can always be added to the two-dimensional angular momentum. We fix it to $C=0$ such that, for $B=0$,
$J$ becomes the standard orbital angular momentum. 
Note that $\Phi(r)$ depends on $B$, and thus is a gauge-invariant quantity, as desired. Altogether, we obtain
\begin{equation}
 J = x p_y - y p_x - qx A_y + q y A_x - q \Phi(r) .
\label{expressionJ}\end{equation}
Concluding, the operator $ J$ as defined above is a gauge-invariant generator of rotations.

We can now determine the form of $J$ in the case of a rotationally invariant magnetic field in the symmetric gauge.
In this gauge, $\vec A$ is purely azimuthal and rotationally symmetric, so $x A_y - y A_x = {\vec r}\times{\vec A}=r A_\theta$
So the magnetic flux is
\begin{equation}
2\pi \Phi(r) = \oint {\vec A}\cdot d{\vec r} = 2\pi r A_\theta
\end{equation}
Therefore, the $\vec A$ and $\Phi$ terms on~\eqref{expressionJ} cancel and we conclude that:
\begin{equation}
 J = x p_y - y p_x .
\end{equation}

In our case we have a homogeneous, and therefore rotationally invariant magnetic field, and we use the symmetric gauge,
so $J$ is as above. This is the operator that we used in the definition of Eq.~\eqref{Eq:Def:Spin}. If instead we worked with a different gauge, such as, e.g., the Landau gauge, $\Phi(r)$ would be the same but the gauge field terms would differ, and we would
have to use the new form of the operator $ J$. Since the result must be gauge invariant, we expect that for states restricted to the LLL it will always hold that:
\begin{equation}
  J = \int \left( \frac{r^2}{2 \ell_B^2}-1 \right) \Psi_{LLL}^\dagger(\mathbf r) \Psi_{LLL}(\mathbf r) d \mathbf r,
\end{equation}
where $\Psi_{LLL}(\mathbf r)$ is the field of the particles composing the FQHE state restricted to the LLL.

\section{Halperin 221 state}
\label{Sec:Halperin221}

We now consider a Halperin 221 state~\cite{Halperin_1983, Girvin_1995}, that describes the quantum Hall effect of two-component bosonic systems, such as particles with an effective spin $\frac 12$ or bilayer setups. 
For simplicity, we consider a bilayer setup and we call the two components with the conventional names A and B; particles on layer A are labelled by a latin index whereas particles on layer B are labelled by a greek index.
The wavefunction that we study reads:
\begin{align}
 \Psi_{221}(z_i,z_\alpha) \propto &\prod_{i<j} (z_i-z_j)^2 \times \prod_{\alpha < \beta} (z_\alpha-z_\beta)^2  \times \nonumber \\ 
 & \times \prod_{i,\alpha} (z_i-z_\alpha) \; e^{- \frac{\sum_i{|z_i|^2}+\sum_\alpha{|z_\alpha|^2}}{4 \ell_B^2}}.
 \label{Eq:Halperin}
\end{align}
This state is amenable to a study employing the plasma analogy~\cite{Girvin_1984, Renn_1994, DeGail_2008, Biddle_2013}.
Our goal is to characterise the spin of its quasiholes and assess that it satisfies a generalised spin-statistics relation. Note that although in some situations the $A$ and $B$ indexes are interpreted as an effective spin-1/2 system, the spin $J$ is completely unrelated to the SU(2) pseudo-spin of the elementary constituents (and indeed we found it also for spinless particles, in the previous section).

In the case of a multi-component state, there are several possibilities concerning the quasiholes; if we think at the quasihole as an approximation for the adiabatic insertion of a flux quantum at $\eta$, the latter can be thread (i) only in the layer A, (ii) only in the layer B, or (iii) in both layers. 
The quasihole wavefunctions that correspond to these situations are:
\begin{subequations}
\begin{align}
\Psi_{A,q}(z_i, z_\alpha, \eta) \propto & \prod_i (z_i-\eta)^q \times \Psi_{221}; \\
\Psi_{B,q}(z_i, z_\alpha, \eta) \propto & \prod_\alpha (z_\alpha-\eta)^q \times \Psi_{221}; \\
\Psi_{AB,q}(z_i, z_\alpha, \eta) \propto & \prod_i (z_i-\eta)^q \prod_\alpha (z_\alpha-\eta)^q \times \Psi_{221}.
 \end{align}
\end{subequations}
With reasonings similar to those presented in Sec.~\ref{SubSubSec:AnalyticalProof} we can compute the spin of $q$ quasiholes of type $AB$, which reads:
\begin{equation}
 J_{q, AB} = - \frac{q^2}{3}+ \frac{2q}{3}.
 \label{Eq:Halperin:Spin}
\end{equation}
Note that this calculation fails for the $\Gamma=A$ quasiholes, because the density profile at the boundary does not shift rigidly when the $q$ quasiholes are inserted, see App.~\ref{App:Boundaries}.

\begin{figure}[t]
 \includegraphics[width=\linewidth]{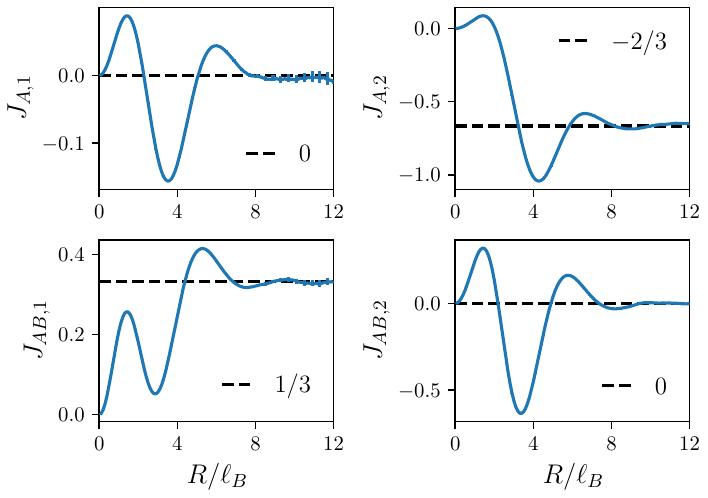}
\caption{
\textbf{Fractional spin of quasiholes of the Halperin 221 state}, $J_{\Gamma,q}(R)$ for $q=1$ ($q=2$), as computed via Monte Carlo sampling and Eq.~\eqref{Eq:Spin_with_cutoff}, for the Halperin 221 states with $N_A = N_B=150$ particles in each layer and for selected values of $(\Gamma,q)$. Note that some properties in Tab.~\ref{Table:Halperin} (e.g. for the exchange $A \leftrightarrow B$) are obtained by a combination of $J_{A,1}$ and $J_{AB,1}$, or by the fact that $J_{A,q}=J_{B,q}$.
The bulk density reads $\bar{\rho} = (2/3)/(2\pi \ell_B^2)$.
Error bars are statistical uncertainties, and data are only shown in a range of cutoff radii $R$ where boundary effects are negligible and Monte Carlo sampling is reliable.
}
\label{Fig:Halperin}
\end{figure}

We compute the spin $J_{\Gamma, q}$ of the quasiholes for $\Gamma = A$ and $AB$ and for $q=1$, $2$, through a Monte Carlo sampling of $|\Psi_{\Gamma,q}|^2$ and by applying the definition in Eq.~\eqref{Eq:Def:Spin}. Our numerical results are presented in Fig.~\ref{Fig:Halperin} and in Appendix~\ref{App:Halperin:Plasma}, where we show the depletion density profile of the quasiholes. 
Similarly to the Laughlin case, the integral $J_{\Gamma,q}(R)$ displays damped oscillations at low values of $R$ and then saturates towards a constant value $J_{\Gamma,q}$, which we find to be compatible with a simple rational number -- as summarized in Table~\ref{Table:Halperin}. To the best of our knowledge, the calculation of the spin of these quasiholes according to existing definitions has not been performed yet. The results for $\Gamma = AB$ coincide with the analytical formula in Eq.~\eqref{Eq:Halperin:Spin}.

Differently from Laughlin's quasiholes, in this case it is not true that $J_{\Gamma, 2} = 4 J_{\Gamma,1}\mod 1$.
Violations of the cluster relation are common in models of non-Abelian anyons, but here we are dealing with Abelian anyons. A similar situation will be discussed in the next section. The violation of the cluster relation is considered acceptable as long as the spin-statistics relation is valid.

\begin{table}[t]
 \begin{tabular}{c || c || c  }
   $\Gamma$ & $J_{\Gamma,1}$ & $J_{\Gamma,2}$  \\
  \hline \hline 
  $A$ & $0$ & $-\frac 23$  \\
  $B$ & $0$ & $ -\frac 23$  \\
  $AB$ & $\frac 13 $ & $ 0$  
 \end{tabular}
 \hfill
 \begin{tabular}{c ||  c }
   $\Gamma  \leftrightarrow \Gamma$ &  $ e^{-  \pi i (J_{\Gamma \Gamma}-2J_{\Gamma,1})}$ \\
  \hline \hline 
  $A \leftrightarrow A$  &
  $e^{ i \frac{2\pi}3}$ \\
  $AB \leftrightarrow AB$  & $e^{ i \frac{2\pi}3 }$ 
 \end{tabular}
 
 \vspace{0.2cm}
 
\begin{tabular}{c || c | c }
$\Gamma \circlearrowleft \Gamma'$ &  $ e^{- 2 \pi i (J_{\Gamma \Gamma'}-J_{\Gamma,1}-J_{\Gamma',1})}$ & Ref.~\cite{Jaworowski_2019}  \\
\hline \hline 
$A \circlearrowleft A$  &
$e^{i \frac{4\pi}3}$ & $e^{ i \frac{4\pi}3}$ \\
$A \circlearrowleft B$ & $e^{i \frac{4\pi}3}$ & $e^{ i \frac{4\pi}3}$ \\
$AB \circlearrowleft AB$  & $e^{ i \frac{4\pi}3 }$ & $e^{ i \frac{4\pi}3}$ 
\end{tabular}
 \caption{\textbf{Fractional properties of the quasiholes of a Halperin 221 state.} (Left): Fractional spin according to the definition in Eq.~\eqref{Eq:Def:Spin}. (Right): Exchange phase. 
 (Bottom): Braiding phase and comparison with the numerical results in Ref.~\cite{Jaworowski_2019}. The spin $J_{\Gamma \Gamma'}$ is the spin of two overlapping quasiholes of type $\Gamma$ and $\Gamma'$. For the sake of clarity: $J_{AA} = J_{A,2}$, $J_{AB} = J_{AB,1}$, and $J_{ABAB} = J_{AB,2}$.}
 \label{Table:Halperin}
\end{table}

When discussing the statistics of the quasiholes of the Halperin 221 state, several possible counter-clockwise exchanges can be envisioned. Here, we consider the following {two} cases: (i) the exchange of two $A$ quasiholes, 
and (ii) the exchange of two $AB$ quasiholes. 
In Ref.~\cite{Jaworowski_2019} the authors compute the \textit{braiding phase} of these two processes and of an additional one, (iii) the encircling of a $A$ quasihole around a $B$ quasihole. 

We compute the exchange phases for the two processes (i) and (ii) using the spin-statistics relation in Eq.~\eqref{Eq:Spin:Statistics} and report the obtained values in Table~\ref{Table:Halperin}. By doubling the obtained phase we obtain the braiding phase for processe (i) and (ii);
for process (iii), we can only compute the braiding phase; our results correspond to the results in Ref.~\cite{Jaworowski_2019}.
We can thus conclude that also in the case of the Halperin 221 state the definition of fractional spin in Eq.~\eqref{Eq:Def:Spin} satisfies the spin-statistics relation.

\section{Some additional observations}\label{Sec:AFI}

We now consider a different model of quasiholes obtained from the rigid shift of the occupation numbers of angular momentum levels; the possibility of characterising exactly their spin $J$ according to Eq.~\eqref{Eq:Def:Spin} will allow us to present some further considerations also for some quasielectron states and to compare with the ``standard'' definition of topological spin.

\subsection{Quasiholes}

We consider a disk geometry where a generic FQHE incompressible fluid with filling factor $\nu$ is realised; in the absence of quasiholes, and in a large-enough system, the occupation number of the lowest angular momentum states is $\nu$ (see Refs.~\cite{MacDonald_1986_A, MacDonald_1986_B} for an early remark on this point, which follows from the assumption that the density-profile of the gas is homogeneous). We introduce the operator $\hat a_l^{(\dagger)}$ that annihilates (creates) a particle in the LLL state with angular momentum $l$, so that for small $l$ we have $\langle \hat a_l^\dagger \hat a_l \rangle_{0} = \nu$. 

We create a quasihole by shifting rigidly the angular momentum occupation numbers by one, and we obtain that the angular momentum occupation number of the state with one quasihole is $ \langle \hat a_l^\dagger \hat a_l \rangle_{1} = \nu$ for $l> 0$, whereas $ \langle \hat a_{0}^\dagger \hat a_{0} \rangle_{1} =0$. More generally, if $q$ quasiholes are created, the angular momentum profile at small $l$ reads:
\begin{equation}
 \langle \hat a_l^\dagger \hat a_l \rangle_{q} = \begin{cases}
       0, & \text{for } l \in \{0, 1, \ldots q-1 \}; \\
       \nu, & \text{for } l \geq q.
 \end{cases}
\label{Eq:AMSON}
\end{equation}

If we consider the Laughlin state $\nu = \frac 1m$, the obtained quasihole does not coincide with the quasihole wavefunction proposed by Laughlin~\cite{Laughlin_1983}; we elaborate on this point in Appendix~\ref{App:Laughlin:AFI}. On the other hand, several properties are discussed in Refs.~\cite{MacDonald_1986_A, MacDonald_1986_B}. It is interesting in the following to compare the properties of the two quasiholes and to see in what they are similar, and in what they are different.

We can re-express our spin definition for the quasihole obtained by inserting $q$ flux quanta~\cite{Sondhi_1992} as: 
\begin{equation}
 J_q(R) = \sum_l \, l \times c_l \, \Big( \langle \hat a_l^\dagger \hat a_l \rangle_{q} - \langle \hat a_l^\dagger \hat a_l \rangle_{0} \Big).
 \label{Eq:Spin:AM}
\end{equation}
where $c_l = \int_0^R |\phi_l|^2 2 \pi r dr$ and $\phi_l$ is the LLL single-particle wavefunction with angular momentum $l$ in the symmetric gauge, see Appendix~\ref{App:Angmom}. 
We consider $R$ large enough to ensure that $c_l \sim 1$ for all values of $l$ such that $\langle \hat a_l^\dagger \hat a_l \rangle_{q} - \langle \hat a_l^\dagger \hat a_l \rangle_{0} $ is significantly different from zero.
In practice, our calculation amounts to computing the sum in~\eqref{Eq:Spin:AM} up to a cutoff angular momentum $\Lambda$ with all $c_l=1$ for $l \leq \Lambda$.

When only one rigid shift is performed, the obtained quasihole has spin $J_1=0$. 
With similar reasoning, for two rigid shifts the spin is $J_2 = -\nu$. 
Again, we violate the generalised cluster relation $J_2 = 4 J_1 \mod 1$, even if we are dealing with Abelian anyons.
With a simple calculation we obtain:
\begin{equation}
 J_{q} = -\frac{q(q-1)}{2} \nu;
 \label{Eq:Spin:AFI:n}
\end{equation}
the spin scales as 
$ \sim q^2$, as desired, but a linear correction is present. Some explicit values are reported in Table~\ref{Table:AFI}.

Using the fractional spins described above, we obtain the correct exchange phase of two $q=1$ quasiholes:
\begin{equation}
 e^{i \Phi_{ex,1}} = e^{-  \pi i (J_2-2J_1 )} = e^{ i \pi \nu}.
\end{equation}
If we consider a Laughlin state with $\nu = 1/m $, we recover the correct value $e^{i \frac{\pi}{m}}$.
If we consider the Halperin 221 state, for which $\nu = \frac 23$, we recover the case of the quasihole of type $AB$. 

We can also compute the exchange phase of two anyons obtained from $q$ rigid shifts:
\begin{equation}
 e^{i \Phi_{ex, q}} = e^{-\pi i (J_{2q}-2 J_q)} =  e^{i  \pi q^2 \nu}.
\end{equation}
Note that the computed exchange phase has the expected scaling as $q^2$. 

\begin{table}[t]
 \begin{tabular}{c || c c c c c | c}
  number of rigid shifts & \phantom{i}$0$\phantom{i} & \phantom{i}$+1$\phantom{i} & \phantom{i}$-1$\phantom{i} & \phantom{i}$+2$\phantom{i} & \phantom{i}$-2$\phantom{i} & \phantom{i}$q$\phantom{i} \\
  \hline
  \phantom{i}$J_q$\phantom{i} & $0$ & $0$ & $- \nu$ & $- \nu$ & $- 3 \nu$ &  \phantom{i}$-\frac{q(q-1)}{2}\nu$\phantom{i} \\
 \end{tabular}
\caption{\textbf{Fractional spin of the quasiparticle obtained by shifting rigidly the occupation numbers of angular momentum states.}}
\label{Table:AFI}
\end{table}

\subsection{Quasielectrons and the ``standard'' definition of topological spin}

An interesting feature of anyons obtained from rigid shifts is the fact that we can also discuss the properties of a specific kind of quasielectron excitations. As originally observed by Laughlin~\cite{Laughlin_1987, MacDonaldQHE}, removing a flux quantum moves the $m=0$ state of the LLL to an excited Landau level state with angular momentum $-1$; if a further flux quantum is removed, the shift continues and the excited-Landau-level state with angular momentum $- 2$ is reached. With this prescription, we can compute the spin $J_q$ of the quasielectrons {by generalising the definition~\eqref{Eq:Spin:AM} to the angular momentum of higher Landau levels}. 
Results are summarised in Table~\ref{Table:AFI}; if we use the label $q>0$ for the quasiholes and $q<0$ for the quasielectrons, the expression for $J_q$ in Eq.~\eqref{Eq:Spin:AFI:n} has a general validity for $q \in \mathbb Z$.
Note that this quasielectron is not confined in the LLL, and therefore it
differs from the one that is typically studied.

On top of the previous calculation concerning the exchange of two quasiholes, we can now compute the exchange of two quasielectrons: 
\begin{equation}
 e^{i \Phi_{ex, -1}} = e^{- i \pi  (J_{-2}- 2 J_{-1})} = e^{ i \pi \nu}.
\end{equation}
The result, that shows that $\Phi_{ex,-1} = \Phi_{ex,1}$, is perfectly consistent with the theory of anyons identifying the quasielectron with the antiparticle of the quasihole.

We can now consider different quasiparticles and the braiding phase associated to the encircling of one quasielectron around one quasihole, and of a quasiparticle composed of $q$ rigid shifts around one composed of $q'$ rigid shifts, with $q,q' \in \mathbb Z$:
\begin{subequations}
 \begin{align}
  e^{i \Phi_{br, 1, -1 }} =& e^{- 2\pi i (J_{0}- J_{-1}-J_{1})} = e^{- 2 \pi i \nu}; \\
  e^{i \Phi_{br, q ,q'}} =& e^{- 2\pi i \left( J_{q+q'} - J_q - J_{q'}\right)} = e^{2 \pi i q q' \nu}.
 \end{align}
\end{subequations}
The result is consistent with the standard expectation for the statistical phase of Abelian anyons upon identification of the quasielectron as the anti-particle of the quasihole.

The characterization of the quasielectrons opens the possibility for the direct comparison with the standard definition of \textit{topological spin} of an anyon. According to this definition, the topological spin of an anyon is associated to the action of the monodromy operator that exchanges an anyon and its anti-anyon when they are in the identity fusion channel. With the previous considerations, we can now compute the spin $s_1$ of the quasihole and $s_{-1}$ of the quasielectron by linking it to half the braiding phase $\Phi_{br, 1, -1}$:
\begin{equation}
 e^{2 \pi i s_1} = e^{2 \pi i s_{-1}} = e^{i \frac{\Phi_{br, 1, -1}}{2}}.
\end{equation}
Usually, the phase $\Phi_{br,1,-1}/2$ is called the statistical angle of the anyon, $\theta$; we obtain here the standard formulation of the spin-statistics relation $s_1 = \theta / 2\pi + \mathbb Z$. 
In our case, $J_1$ does not coincide with $s_1$: 
\begin{equation}
 s_1 = s_{-1}  = \frac{J_1+J_{-1}}{2} ;
\end{equation}
for this reason the authors of Ref.~\cite{Einarsson_1995} propose to speak of a generalised spin-statistics relation 
\begin{equation}
\frac {\theta}{2 \pi}  = \frac{J_1+J_{-1}}{2} + \mathbb Z = - \frac{\nu}{2}+ \mathbb Z.
\end{equation}

We can generalise this discussion to the case of anyons composed of $q$ flux quanta:
\begin{equation}
 s_q = s_{-q} = \frac{J_q+J_{-q}}{2} .
\end{equation}
Our fractional spin $J_q$ allows to reconstruct the spin $s_q$ according to the standard definition; as it was already noticed, $s_q$ only depends on the even part (with respect to $q$) of $J_q$, which leaves the value of the odd part unconstrained and thus explains the difference between $J_q$ and $s_q$.
The statistical parameter reads:
\begin{equation}
 \theta_q = \frac{J_q+J_{-q}}{2} + \mathbb Z = - \frac{q^2 \nu}{2}+ \mathbb Z = q^2 \theta_1+ \mathbb Z.
\end{equation}

The comparison between the data obtained on the Laughlin state in Sec.~\ref{Sec:GeneralisedSpinStatistics} and the angular-momentum rigid shift at $\nu = \frac 1m$ shows that quasiholes with the same statistical properties are characterised by the same $s_q$ but by different $J_q$; it thus appears that $s_q$ is an  invariant, whereas $J_q$ is not. 
In fact, any spin of the form $J_q = - \frac{q^2}{2m}+ q c$, where $c$ is a constant, would give the correct $s_q$. 

\section{Adiabatic rotation of quasiholes}
\label{Sec:AdiabaticRotation}

The results in the previous sections, and in particular the systematic spin-statistics relation~\eqref{Eq:Spin:Statistics} satisfied by our spin, demand for a general interpretation. In this section we present a theoretical framework that explains their origin; this discussion is a generalization of the analysis in Refs.~\cite{Umucalilar_2018, Macaluso_2019, Macaluso_2020}, which was limited to the derivation of the braiding phase.   

\subsection{General theoretical framework}

\begin{figure*}[t]
\includegraphics{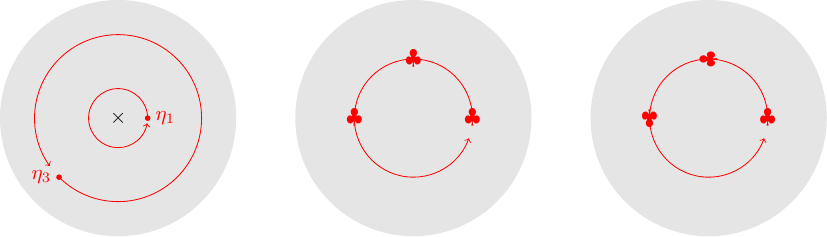}
\caption{(Left) Sketch of a FQHE state defined on a disk with two quasiholes located at $\eta_\mu$, $\mu=1,2$. The red lines indicate the trajectories of the quasiholes, which are rigidly rotated of an angle $\theta_f = 2 \pi$ at fixed distance $|\eta_i|$ from the center of the disk ($\times$).
(Center and right)
Two different ways of moving quasiparticles: without self-rotation, as in the left panel, or with self-rotation, as in the right panel. The former approach is the standard one, while the latter is the one considered in this article; in order to study it, we keep the quasiparticles fixed in the reference frame that rotates at angular frequency $\vartheta_f / T$.
}
\label{Fig:RigidRotation}\label{Fig:TwoRotationSchemes}

\end{figure*}

We consider a circularly-symmetric FQHE state composed of $N$ quantum particles with complex coordinates $z=x+iy$ in the LLL; they can be bosons or fermions and we consider in both cases a positive charge $+e$ ($e>0$ is the fundamental electric charge); this is the convention used also in Ref.~\cite{Einarsson_1995}. 
We consider some external pinning potentials that are placed at the positions $\eta_\mu \in \mathbb C$ far from the boundary and such that $|\eta_\mu-\eta_\lambda|$ is much larger than any physical length characterising the problem; we assume that one quasiparticle binds to each pinning potential. 

In order to reveal the rotational properties of the particles, and hence their spin, we propose to move the pinning-potential coordinates of an angle $\vartheta_f$ around the center of the disk in time $T$ in a counter-clockwise fashion, see Fig.~\ref{Fig:RigidRotation} (left):
    \begin{equation}
    \eta_\mu (t) = \eta_i (0) e^{+i \vartheta(t)},
    \qquad \vartheta(t) = \frac {t}{T} \vartheta_f, \quad t \in [0, T] .
    \label{Eq:Anyon:Rotation}
    \end{equation}
As it was already observed~\cite{read2008quasiparticle, Read_2009}, this rotation does not necessarily rotate the quasiparticles around themselves.
However, rotational properties can be studied once we define the problem in the reference frame $R_2$ that is co-rotating with the pinning potentials at angular frequency $\vartheta_f/T$. If the quasiparticle is held fixed in this reference frame, then it rotates around itself in the laboratory reference frame.
Fig.~\ref{Fig:TwoRotationSchemes}~(center and right) provides a pictorial sketch of this intuition.
The study of the Berry phase due to the quasiparticles obtained within this scheme will allow us to define the quasiparticle spin.

We place ourselves in $R_2$ and write the generator of the time-evolution in this reference frame.
The FQHE Hamiltonian $\hat H_{\rm FQHE}$ written in the symmetric gauge is rotationally invariant: it is thus the same in the laboratory and in $R_2$; 
if a confining potential is present, we assume for simplicity that it is rotationally invariant.
Concerning the pinning potentials,
we have to assume the potential responsible for the pinning, $\hat V_{\rm pinning} (z_i, \eta_\mu(t))$, is not rotationally invariant because of an infinitesimal rotational asymmetry, so that no confusion between the two sketches in Fig.~\ref{Fig:TwoRotationSchemes}~(center and right) is possible.
Under these assumptions, the generator of the time evolution in $R_2$ in the time interval $[0,T]$ reads~
\begin{equation}
 \hat H_2 =
 \hat H_{\rm FQHE} + \hat V_{\rm pinning}(\, z_i, \eta_\mu(0) \,)
 - \frac{\vartheta_f}{T} \hat L_z,
 \label{Eq:Ham:R2}
\end{equation}
which is manifestly time-independent. 
The last term, proportional to the rotation frequency, represents (i) a centrifugal potential due to the fact that $R_2$ is a non-inertial reference frame, and (ii) the scalar potential associated to the radial electric field that appears in a rotating reference frame when there is a homogeneous magnetic field in the laboratory -- see Appendix~\ref{Appendix:RotatingFrame}. 
Since we are interested in an adiabatic process with $T \to \infty$, and since the bulk of the FQHE system is gapped, we will treat this term perturbatively.
A problem remains because our system has gapless edge modes. We can safely disregard this issue because we are interested in bulk properties, and the dynamics in the bulk effectively decouples from the edges. Thus, provided we do not focus on the edges, we can employ the adiabatic theorem and consider the centrifugal potential in Eq.~\eqref{Eq:Ham:R2} as a small perturbation, and use time-dependent perturbation theory restricted to the ground-state subspace. 

To describe the dynamics in $R_2$, we consider an initial state $\ket{\Psi_2(0)}$ belonging to the $k$-fold degenerate ground-state subspace $ \mathcal{H}_{E_0}$ of $\hat H_{\rm FQHE}+ \hat V_{\rm pinning}$, spanned by the basis $\lbrace\ket{\psi_{\alpha}}\rbrace_{\alpha=1,\dots,k}$, with energy $E_0$ and $\bra{\psi_{\alpha}}\psi_{\beta} \rangle = \delta_{\alpha\beta}$.
If the dynamics is slow enough, we can use the adiabatic theorem to state that the dynamics is restricted to $\mathcal{H}_{E_0}$, and make the following ansatz:
\begin{equation}
\ket{\Psi_2(t)} =
e^{- i E_0 t } \, \sum_{\alpha = 1}^k \gamma_{\alpha}(t) \ket{\psi_{\alpha}},
\quad  \gamma_\alpha(0) = \bra{\psi_{\alpha}}  \Psi_2(0)\rangle. \label{Eq:State:2:Ansatz}
\end{equation}
By applying the Schr\"odinger equation, we recover the time-evolution equation of the $\gamma_\alpha$s:
\begin{subequations}
\begin{align}
 i  \frac{ d \gamma_{\alpha}(t)}{ d t} =&
 -\frac{\vartheta_f}{T}\sum_{\beta=1}^k 
 \mathcal{L}_{\alpha\beta}\,  \gamma_{\beta} (t), 
 \\
 \mathcal{L}_{\alpha\beta} =& \langle\psi_{\alpha} |  \hat{L}_z  | \psi_{\beta} \rangle.
 \label{Eq:Diff:gamma}
\end{align}
\end{subequations}
The solution reads
$
 \ket{\Psi_2(T)} = 
  e^{- i E_0 T }\,
 e^{i \vartheta_f \mathcal L }
 \ket{\Psi_2(0)},
$
where $\exp \left[i \vartheta_f \mathcal L  \right]$ is the matrix exponential of $i \vartheta_f\mathcal L$. 

In order to find the state $\ket{\Psi_1(T)}$ in the laboratory frame, we need to rotate back $\ket{\Psi_2(T)}$ by an angle $\vartheta_f$:
\begin{align}
\ket{\Psi_1(T)}
&=
e^{- i \vartheta_f \hat{ L}_z }  \ket{\Psi_2(T)} = \nonumber \\
&=
e^{- i E_0 T }\,
e^{- i \vartheta_f \hat{ L}_z }\,
e^{i \vartheta_f \mathcal L }  \ket{\Psi_2(0)}.
\label{Eq:Psi1}
\end{align}
Note that $\ket{\Psi_1(0)} = \ket{\Psi_2(0)}$.
The state in Eq.~\eqref{Eq:Psi1} is the exact result for the rotation by an angle $\vartheta_f$ performed including the self-rotation of the quasiholes.
We recognize a dynamical phase proportional to $T$, that is unessential to our discussion and therefore neglected from now on.
Moreover, when $\vartheta_f = 2 \pi $ a further simplification is possible because $e^{ 2 \pi i\hat L_z } = 1$. 
In this case, $\mathcal L$ encodes the full geometric contribution to the time evolution, which for a non-degenerate state reads:
\begin{equation}
 \ket{\Psi_1(T)} \, \propto  \,
 e^{2 \pi i \, \langle \Psi_1(0) | \hat L_z | \Psi_1(0) \rangle} \, \ket{\Psi_1(0)}.
\end{equation}
The study of the geometric phase associated to $\langle \Psi_1(0) | \hat L_z  | \Psi_1(0) \rangle$ in various configurations of the pinning potentials will give us information about the spin and statistics of the quasiholes.

\subsection{No quasiparticles as a regularisation}
\label{Sec:ZeroQH}

In this article we are interested in understanding how the quasiparticles, that are localised objects, affect the geometric phase. 
We thus take the state $\ket{\psi}$ without quasiparticles, for which the geometric phase is trivial, as a reference for any further calculation: it will act as a regularising background. 
When studying the state $\ket{\varphi}$ characterised by the presence of quasiparticles, we will thus focus on the quantity:
\begin{equation}
 \Delta L_{z, \varphi} = \bra{\varphi} \hat L_z \ket{\varphi}- \bra{\psi} \hat L_z \ket{\psi}.
 \label{Eq:DeltaLz}
\end{equation}
and identify the contribution to the geometric phase that can be uniquely ascribed to the quasiparticle.

\subsection{Spin of the quasiparticles}
\label{Sec:OneQH}

\begin{figure}[t]
\centering 
\includegraphics{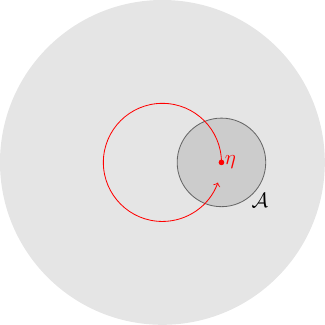}
\caption{FQHE state with one quasiparticle located at $\eta$. Red: Adiabatic motion of one quasiparticle around a circumference with radius $|\eta|$. Black: region $\mathcal A$ used to perform the integral in Eq.~\eqref{Eq:1qh} and to avoid boundary effects.}
\label{Fig:OneQuasihole}
\end{figure}

We consider the non-degenerate quantum state $\ket{\psi_\eta}$ with one quasiparticle located at $\eta$, see Fig.~\ref{Fig:OneQuasihole}; our first goal is to compute $\Delta L_{z,\psi_\eta}$ defined in Eq.~\eqref{Eq:DeltaLz}. 
We introduce the radial density profile of $\ket{\psi}$, which we call $\rho(|z|)$, 
and the density profile $\rho_\eta(z)$ of $\ket{\psi_\eta}$. 
By using the properties of the angular momentum operator $\hat L_z$ in the LLL recalled in Appendix~\ref{App:Angmom}, we obtain:
\begin{equation}
 \Delta L_{z,\psi_\eta} =
  \int \left(\frac{|z|^2}{2  \ell_B^2}-1 \right) 
  \big[ \rho_\eta(z) - \rho(|z|) \big] d^2z .
 \label{Eq:1qh}
\end{equation}
The quasiparticles are localised objects, thus the two density profiles differ significantly only in two regions of space: (i) around the quasiparticle located at $\eta$, and (ii) at the boundary.
Since we want to characterise the properties of the quasiparticle, we are not interested in the boundary contribution, which will be from now on neglected. 
In a numerical calculation, the boundary can be discarded by restricting the integral in~\eqref{Eq:1qh} to a region $\mathcal A$ with radius much smaller than the droplet size and centered around $\eta$, see for instance Fig.~\ref{Fig:OneQuasihole}.

For the sake of simplicity, we now assume to have an infinite system and work with the depletion density $d_\eta(z) = \rho_\eta(z) - \bar{\rho}$ induced by the quasiparticle, where we use the fact that the density profile of $\ket{\psi}$ is uniform and equal to $\bar{\rho}$, in the absence of boundary.
The depletion density goes to zero for $|z|\to \infty$ and, in the case of a circularly symmetric quasiparticle, it
only depends on the distance $|z - \eta|$. 

We shift the origin to $\eta$ introducing the new variable $z' = z-\eta$ and reformulate one part of the integral in Eq.~\eqref{Eq:1qh} as follows:
\begin{equation}
  \int \frac{|z|^2}{2 \ell_B^2} d_\eta(|z-\eta|) d^2z = \int \frac{|z'+\eta|^2}{2 \ell_B^2} d_0(|z'|) d^2z'.
\end{equation}
where $d_0(z)$ is the depletion density of the quasiparticle when it is placed at the origin.
This integral can be simplified by using the polar representation $z'=r' e^{i \phi'}$  and expanding it into three parts:
\begin{equation}
 \int \frac{r'^2+2  r'|\eta| \cos \phi'+|\eta|^2}{2 \ell_B^2} d_0(r') \, r' dr' d \phi'.
\end{equation}
By performing the integral over $\phi'$ we can disregard the term proportional to $|\eta|$;
we finally obtain:
\begin{equation}
 \Delta L_{z,\psi_\eta}= 2 \pi 
 \int_0^\infty \left(\frac{r^2}{2 \ell_B^2}-1 \right)d_0(r) r 
 dr  +  \frac{Q_1}{e} \frac{|\eta|^2}{2 \ell_B^2} .
\label{Eq:1qh:Bphase}
 \end{equation}
where the charge of one quasiparticle reads
\begin{equation}
 Q_1 = e \int d_0(|z|)d^2z = 2 \pi e \int_0^\infty d_0(r) r dr .
\end{equation}

The result in Eq.~\eqref{Eq:1qh:Bphase} includes two terms, one of which depends on the quasiparticle position via $|\eta|^2$. It is a well-known result in the theory of FQHE that the geometric phase picked up by a quasiparticle moving in the plane should contain a term proportional to the area enclosed by the rotation, which in our case is $A_\eta =\pi |\eta|^2$. This phase reads $\exp \left[ + i \frac{Q_1}{e}\frac{A_\eta }{\ell_B^2}  \right]$.
Note that $A_\eta / \ell_B^2 = 2 \pi \Phi_\eta / \Phi_0$ with $\Phi_\eta$ the magnetic flux enclosed in the circle with radius $|\eta|$ and $\Phi_0 $ the magnetic flux quantum, so that the phase reads $\exp \left[ + i Q_1 \Phi_\eta  \right]$; 
this is the Aharonov-Bohm phase picked up by the charge, a term which is expected from standard textbook considerations and is not linked to the self-rotation of the quasiparticle, but only to its translation in the plane.

Concerning the rest of the geometric phase, the term:
\begin{equation}
 \ell_{z,1} =  2\pi 
 \int_0^\infty \left(\frac{r^2}{2\ell_B^2}-1 \right)d_0(r) r 
 dr 
\end{equation}
describes the difference in angular momentum due to one quasiparticle and it appears in the geometric phase because during the adiabatic motion the quasiparticle has rotated around itself.
This term does not appear in the standard discussions on the geometric phase associated to the motion of a single quasiparticle on a disk because one typically studies a purely translational motion. The peculiarity of the approach that we are suggesting is that the quasihole rotates while moving, and this explains \textit{a posteriori} the appearance of the new term.

In summary, the geometric phase associated to the motion of one quasiparticle that we just computed,
\begin{equation}
 e^{i \phi_\eta} = e^{2 \pi i \ell_{z,1}} e^{i  Q_1 \Phi_\eta},
 \label{Eq:Gphase:1qh}
\end{equation}
has two contributions which are amenable to simple physical interpretation.
Our scheme has the remarkable property that in the limit $|\eta|\to 0$ the system still picks up a geometric phase, because the quasiparticle rotates around itself. 
We thus propose to interpret this term as the \textit{spin of the quasiparticle}:
\begin{equation}
 J_1 =  \ell_{z,1} 
 \label{Eq:Spin}
\end{equation}
so that a counter-clockwise rotation of the quasiparticle by $2 \pi$ is associated to the phase $e^{+i 2 \pi J_1}$. This is the definition that we have presented at the beginning of the article in Eq.~\eqref{Eq:Def:Spin}.

\subsection{Statistics of the quasiparticles}
\label{Sec:TwoQH}

\begin{figure}[t]
\centering
\includegraphics{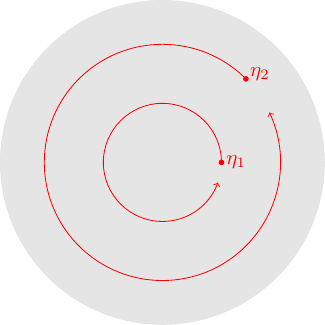}
\caption{LLL state with two quasiparticles located at $\eta_1$ and $\eta_2$. Red: Adiabatic motion of one quasihole around a circumference with radius $|\eta_\mu|$.}
\label{Fig:TwoQuasiholes}
\end{figure}

We now move to the discussion of the spin-statistics relation and consider the state $\ket{\psi_{\eta_1, \eta_2}}$ with two quasiparticles at positions $\eta_\mu$, and by convention we set $|\eta_2|\geq|\eta_1|$. 
We further assume that the state is not degenerate; this assumption does not preclude the study of quasiparticles with non-Abelian statistics, because we are simply assuming that they are in a well-defined fusion channel.

We introduce the density profile $\rho_{\eta_1, \eta_2}(z)$ and address the geometric phase picked up by the system when the two quasiparticles are rigidly rotated at fixed distance from the center, as sketched in Fig.~\ref{Fig:TwoQuasiholes}. To this goal, we compute:
\begin{equation}
 \Delta L_{z,\psi_{\eta_1, \eta_2}} = \int \left(\frac{|z|^2}{2 \ell_B^2}-1 \right) \Big( \rho_{\eta_1, \eta_2}(z) - \rho(|z|) \Big) d^2z .
 \label{Eq:2qh}
\end{equation}
This time, the quantity $\rho_{\eta_1, \eta_2}(z) - \rho(|z|)$ differs from zero in three regions of space: (i) around the quasiparticle $\eta_1$, (ii) around the quasiparticle $\eta_2$, and (iii) at the boundary. We neglect this latter contribution as our focus is on quasiparticles and we effectively consider an infinite system. 

In general, we expect the geometric phase to have a dependence on the position of the two quasiholes, $\varphi_{\eta_1, \eta_2}$, and in particular on the distance $|\eta_1-\eta_2|$.
The two limiting cases $|\eta_1-\eta_2|\gg l_B$ and $|\eta_1-\eta_2|=0$ are easy to treat.
If the two quasiholes are put sufficiently far apart, the approximate equality
\begin{equation}
 \rho_{\eta_1, \eta_2}(z) - \rho(|z|) \approx d_{\eta_1}(z)+d_{\eta_2}(z)
\end{equation}
allows us to use the results of the previous section and to write
\begin{equation}
 e^{i \phi_{ \eta_1, \eta_2}} \approx e^{i \phi_{\eta_1}} e^{i \phi_{ \eta_2}} \quad \text{for } |\eta_1-\eta_2| \gg l_B.
 \label{Eq:GPhase:2qh:Far}
\end{equation}
where $e^{i \phi_{\eta_1}}$ is defined in Eq.~\eqref{Eq:Gphase:1qh}.

The situation of two overlapping quasiholes is studied by putting both of them at the origin and introducing the associated depletion density:
\begin{equation}
 d_{0, 0}(z) = \rho_{0, 0}(z) - \bar{\rho}.
\end{equation}
With steps similar to those of the previous section, we obtain:
\begin{equation}
 e^{i \phi_{\eta,\eta}} = e^{ 2 \pi i \ell_{z,2}} e^{+i \frac{Q_2}{e} \frac{ A_\eta}{\ell_B^2}} \quad \text{for } \eta_1=\eta_2=\eta,
 \label{Eq:GPhase:2qh:OnTop}
\end{equation}
where:
\begin{subequations}
\begin{align}
 Q_{2} =& 2 \pi e \int_{0}^\infty d_{0,0}(r) rdr;\\
 \ell_{z,2} =& 2 \pi \int_{0}^\infty \left( \frac{|z|^2}{2 \ell_B^2}-1\right) d_{0,0}(z) rdr. 
\end{align}
\end{subequations}
Note that $Q_{2} = 2 Q_{1}$ whereas in general $\ell_{z,2} \neq 2 \ell_{z,1}$. 

It easy to identify in Eq.~\eqref{Eq:GPhase:2qh:OnTop} a term that is an Aharonov-Bohm phase; in fact, it is possible to give an expression of the Aharonov-Bohm phase for a generic set of positions of the two quasiholes, $ \eta_1$ and $\eta_2$,
\begin{equation}
  e^{{ +}i Q_1 \Phi_{\eta_1}} \, 
  e^{{ +}i Q_1 \Phi_{\eta_2}},
\end{equation}
and to observe that the two expressions appearing in Eqs.~\eqref{Eq:GPhase:2qh:Far} and~\eqref{Eq:GPhase:2qh:OnTop} are just two specific cases.

The remaining term $\Delta\phi$ is also a function of the position of the quasiholes and of their distance. 
The two limiting values are known:
\begin{equation}
 \Delta \phi = {2 \pi}  \times \begin{cases}
                2 \ell_{z, 1}, & |\eta_1-\eta_2| \gg l_B \\
                \ell_{z, 2}, & |\eta_1-\eta_2|=0.
               \end{cases}
\end{equation}
The fact that $\Delta \varphi$ depends on $|\eta_1-\eta_2|$ can be explained once we observe that
when the two quasiparticles are far apart they are braiding; instead, when the two quasiparticles are on top of each other, they are not braiding.
We thus pose that the braiding phase is the only responsible for this phase difference and write that:
\begin{equation}
 e^{i\Phi_{br}} = e^{2 \pi i \left(2 \ell_{z,1}- \ell_{z,2} \right) }.
\label{Eq:BraidingPHase}
 \end{equation}
If we parallel the definition in Eq.~\eqref{Eq:Spin} and define the spin of the quasiholes composed of two elementary quasiholes as:
\begin{equation}
 J_2 =  {+}\ell_{z,2}
\end{equation}
we obtain:
\begin{equation}
 e^{i \Phi_{br}} = e^{{-}2 \pi i(J_2 - 2 J_1 )}.
\end{equation}

This result can be read as a spin-statistics relation for the quasiparticle. Although the discussion has been carried out assuming that the two quasiparticles are of the same kind, it is not difficult to repeat the derivation for a quantum state composed of two different quasiholes, say $a$ and $b$, which compose into a quasihole of type $c$. In that case:
\begin{equation}
 e^{i  \Phi_{br}} = e^{- 2\pi i \left( J_c - J_a - J_b\right)}
\end{equation}
which is the most general form in which the spin-statistics theorem is typically enunciated for an anyon model (see Eq.~\eqref{Eq:Spin:Statistics:2}).

In order to consider the exchange phase of two quasiparticles of type $a$, we simply divide by two the phase and obtain the spin-statistics relation in Eq.~\eqref{Eq:Spin:Statistics}:
\begin{equation}
 e^{i \Phi_{ex}} = e^{-i \pi \left( J_c -  2J_a \right)}
\end{equation}
This concludes our discussion and shows in very general terms that our spin is compatible with a generalised spin-statistics theorem. The several numerical results reported in Secs.~\ref{Sec:GeneralisedSpinStatistics}, \ref{Sec:Halperin221} and~\ref{Sec:AFI} are not a coincidence, but originate from deep reasons related to the rotational properties of the quasiparticles.

\section{Conclusions}
\label{Sec:Conclusions} 

The theory of anyons is deeply rooted in the spin-statistics theorem. Quasiparticle excitations of incompressible FQHE states have been widely studied and several fractional properties have been pointed out, notably their charge and their statistics. 
It is a reasonable expectation that these particles should also have a fractional spin; yet, this was assessed only for systems defined on a spherical surface. 
In this article we have proposed a measurable spin for such quasiparticles and we have shown with numerical and analytical methods that it satisfies a generalised spin-statistics relation. 

Two important remarks are in order. The first one is that the fractional spin discussed in the article is not related to the SU(2) spin-1/2 that characterizes electrons. We are here speaking of an emergent property of a many-body quantum state, and indeed we also applied the formula to model wavefunctions for spinless particles. The second remark is that even if in all examples discussed so far we have only considered Abelian anyons, nothing precludes the application of our formula to non-Abelian anyons~\cite{Macaluso_2019}. Once a given fusion channel is chosen, Eq.~\eqref{Eq:Spin:Statistics} for the spin-statistics relation applies.

The possible implications of our result are copious.
From an experimental viewpoint, it opens a new path towards the assessment of the anyonic properties of the FQHE quasiparticles.
Recent experimental advances in the realisation of FQHE states with photonic and atomic gases~\cite{Tai_2017, Aidelsburger_2018, Cooper_2019, Chalopin_2020, Jamadi_2020, Clark_2020, Rahmani_2020}, also by means of synthetic dimensions~\cite{Mancini_2015,Stuhl_2015, Fabre_2021}, make it possible to envision the study of strongly-correlated liquids using observables that are radically different from those that are standard in solid-state setups;
in particular, a characterisation of the system in real space is possible~\cite{Raciunas_2018, Umucalilar_2018_B, Rosson_2019, Repellin_2019, Repellin_2020, delasheras_2020, Grass_2020, Baldelli_2021, Cian_2021, wang2021measurable}.   
Our work fits well in this endeavour and shows that the study of the charge and of the quadrupole moment of the quasiparticle is sufficient to assess its fractional spin. By also studying small clusters of quasiparticles, it is then possible to infer the fractional statistics without using any interferometry~\cite{Umucalilar_2018, Macaluso_2019, Macaluso_2020}.

Additionally, one could look at our result from a different perspective, and observe that we have shown that the quasiparticles of the FQHE fractionalise the \textit{canonical} angular momentum of the gas (the gauge-invariant meaning of this observable has been discussed). Since it does not coincide with the mechanical angular momentum, it is not related to the currents flowing in the system nor to its magnetisation~\cite{Platzman_1986}. Yet, it is an intriguing perspective to speculate on the possible experimental consequences of this fact, which could be observed, for instance, probing the rotational properties of the gas.

On the theoretical side, the possibility of defining the fractional spin on a non-curved surface motivates further investigations in the context of FQHE liquids defined on a cylinder, a setup that is amenable to a natural matrix-product-state description~\cite{Zaletel_2012, Estienne_2013}. 
It would be extremely interesting to apply our spin definition to several quasiparticles that have been characterised thanks to matrix-product states.
As a first example, let us mention the quasielectron of the Laughlin state~\cite{Kjonsberg_1999, Kjall_2018}; a  characterisation of its fractional spin would allow us to assess the topological spin of the quasihole according to the standard definition, $S = \frac{1}{4 \pi} (J_{qh}+ J_{qe})$.
As a second example, let us mention the study of the quasiparticles of the $\nu = \frac{12}{5}$ and $\frac{13}{5}$ states that are expected to be a good candidate for Fibonacci anyons~\cite{Mong_2017}. Since several quasiparticles excitations with Abelian or non-Abelian statistics can be distinguished by their quadrupole moment, it is interesting to link this observation to our definition of fractional spin. 
A first step in this direction has already been performed in Ref.~\cite{Macaluso_2019}, while discussing the link between the quadrupole moment and the braiding phase of the non-Abelian anyons of the Moore-Read wavefunction.

In some situations, we have been able to compute the spin of the quasiparticle analytically; for the quasihole of the Laughlin state, for instance, we have shown that $J_{m,q} = - \frac{q^2}{2m}+ \frac{q}{2}$. This opens the interesting perspective of generalising this formula to other FQHE quasiparticles and, ideally, to other quasiparticles that have fractional spin.

Finally, we would like to stress that a microscopic understanding of the proposed spin-statistics relation is still missing. 
According to our discussion, the statistics of the quasiparticles is written in the quadrupole moment of clusters of quasiholes,
but, as it was early noticed in Ref.~\cite{Sondhi_1992}, the quadrupole moment of the fractional charge depends on the specific form of the model, on the particular form of the interparticle interaction, and no quantisation is expected. 
This effect was really tangible when we have compared the quasiholes of a Laughlin state with the anyons obtained by angular-momentum rigid shift. Although the statistical properties of these quasiparticles are the same, their spins $J$ are different.
This calls for a microscopic explanation, that is left for future work.

\acknowledgments

We warmly acknowledge enlightening discussions with N.~Baldelli, V.~Cheianov, B.~Estienne, M.~Goerbig, T.~Gra\ss, G.~La Rocca, S.~Ouvry, N.~Regnault, L.~Rosso and J.~Slingerland.
We are indebted with B.~Estienne for suggesting us to investigate the problem of topological spin and for pointing out to us Ref.~\cite{read2008quasiparticle}.
This work follows a series of studies performed with I.~Carusotto and O.~Umucal$\i$lar: we thank them for uncountable discussions on the topic in the last years. We thank B.~Jaworowski for sharing  with us data from Ref.~\cite{Jaworowski_2019}.
A.~P.~P. thanks LPTMS for hospitality.
T.C.~acknowledges funding by the Agence Nationale de la Recherche (EELS project, ANR-18-CE47-0004);
A.B.~and L.M.~acknowledge funding from LabEx PALM (ANR-10-LABX-0039-PALM).
All numerical simulations have been performed on the PSMN cluster of the ENS of Lyon, and we make the numerical density profiles publicly available \cite{DATASET_2021}.


\appendix

\section{Lowest Landau level and angular momentum in the symmetric gauge}
\label{App:Angmom}

For the sake of clarity, we report explicitly a few useful formulas for the angular momentum ${L}_z = x p_y- y p_x$ in the LLL.
We start from the LLL normalized single-particle wavefunctions written in the symmetric gauge,
\begin{equation}
 \phi_m(z) = \frac{i^m}{\sqrt{2 \pi \ell_B^2 m!}} \left( \frac{z}{\sqrt 2 \ell_B} \right)^m e^{- |z|^2 / 4 \ell_B^2}, \quad m \in \mathbb N.
\end{equation}
Using the polar-coordinate parametrisation of the complex plane, $z = r e^{i \theta}$, we can write ${L}_z = - i \partial_\theta$, from which we obtain $ L_z \phi_m = m \phi_m$. We now use the following fundamental integral, which is simply derived from the properties of the Euler's $\Gamma(x)$ function:
\begin{equation}
 2 \pi \int_0^\infty |\phi_m|^2 \frac{r^2}{2 \ell_B^2}  r \,dr = m+1
\end{equation}
to deduce that:
\begin{equation}
 \int \phi_m^* L_z \phi_m r dr d \theta = m = \int |\phi_m|^2 \left( \frac{r^2}{2 \ell_B^2}-1\right) r dr d \theta.
\end{equation}
This expression explains the connection between the angular momentum operator restricted to the LLL and the spatial profile of the wavefunction.
In a second-quantisation description of the problem, we introduce the LLL field,
\begin{equation}
 \Psi_{LLL}(z) = \sum_{m=0}^\infty \phi_m(z)  a_m,
\end{equation}
in order to write the angular momentum operator as
\begin{equation}
 L_z = \sum_m m  a^\dagger_m  a_m =  \int  \Psi_{LLL}^\dagger \left( \frac{|z|^2}{2 \ell_B^2} - 1\right)  \Psi_{LLL} d^2z.
\end{equation}
This expression is used in the definition of the fractional spin -- Eq.~\eqref{Eq:Def:Spin}.

\section{Boundary properties of density profiles}
\label{App:Boundaries}

\begin{figure}[t]
\includegraphics[width=0.98\linewidth]{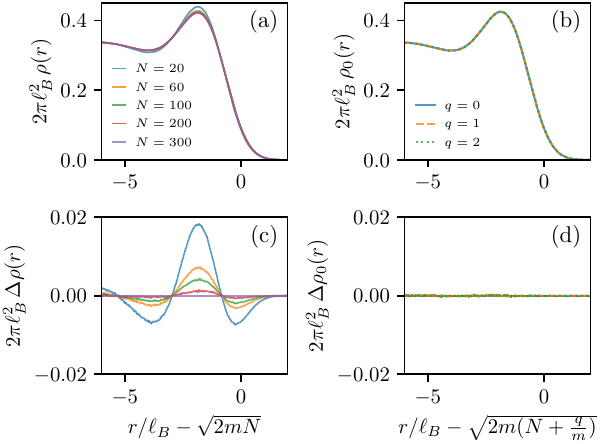}
 \caption{Laughlin density profile at the boundary, obtained by Monte Carlo sampling:
(a) Density profiles for different sizes $N$, in the absence of quasiholes ($q=0$);
(c) same data as in (a), after subtracting the curve for $N=300$;
(b) density profiles for different numbers $q$ of quasiholes in the origin (for $N=100$);
(d) same data as in (b), after subtracting the curve for $q=0$.
}
\label{Fig:Laughlin:Boundary}
\end{figure}

In Section \ref{Laughlin:Proof}, we proved the expression in Eq.~\eqref{Eq:Laughlin:J:Proof} for the spin of the Laughlin quasiholes; a similar result is presented in Eq.~\eqref{Eq:Halperin:Spin} for the $\Gamma=AB$ quasiholes of the Halperin 221 state. 
A crucial assumption of this proof is that the density profile shifts in a rigid way upon increasing either the system size $N$ or the number $q$ of quasiholes in the origin. Here we verify whether these assumptions by means of Monte Carlo simulations for the Laughlin and Halperin 221 states.

\begin{figure}[t]
\includegraphics[width=0.98\linewidth]{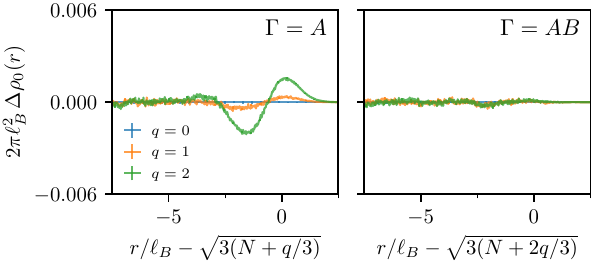}
 \caption{
(a) Halperin 221 density profile at the boundaries for different numbers $q$ of $\Gamma=A$ quasiholes in the origin (for $N=100$), after subtracting the $q=0$ curve;
(b) same as in (a), for $\Gamma=AB$.
}
\label{Fig:Halperin:Boundary}
\end{figure}

For the Laughlin state, panels (a) and (c) of Fig.~\ref{Fig:Laughlin:Boundary} show the density profile at the boundary of the system, for different values of $N$. This profile is not fully universal (as a function of the shifted radius $r/\ell_B-\sqrt{2mN}$), due to  non-universal finite-size corrections that disappear for large systems and that thus have not been included in the proof.
In the same figure, panels (b) and (d) show that the density profile at the boundary is fully universal when adding $q$ quasiholes in the origin (after a shift of the radius by $\ell_B \sqrt{2m(N+q/m)}$).
This shows that the assumptions on the rigid shift of the density profile are valid.

For the Halperin state, a similar finite-$N$ correction is present [not shown], while the evolution of the density profile at the boundary upon increasing $q$ shows a peculiar dependence on $\Gamma$. As shown in Fig.~\ref{Fig:Halperin:Boundary}, only in the $\Gamma=AB$ case we observe a rigid shift of the density profile, while for $\Gamma=A$ the shift is not rigid. For this reason, we are able to derive Eq.~\eqref{Eq:Halperin:Spin} for the $\Gamma=AB$ quasiholes, but the procedure fails when it is applied to $\Gamma=A$.

\section{Plasma analogy for the Halperin 221 state}
\label{App:Halperin:Plasma}

\begin{figure}[t]
\includegraphics[width= 0.98\linewidth]{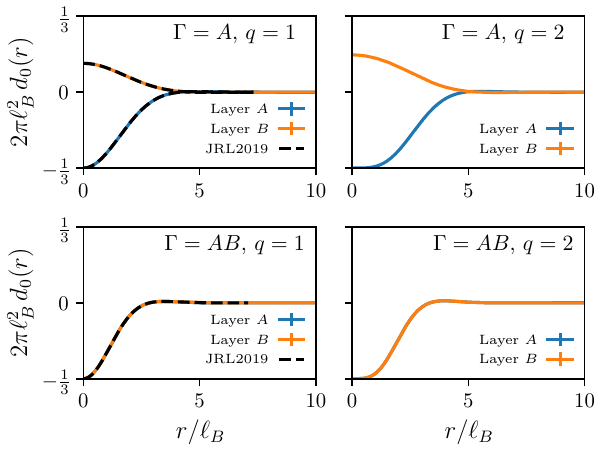}
 \caption{Depletion density profile on the A and B layers, obtained by Monte Carlo sampling of the Halperin 221 state with $\Gamma=A,AB$ and $q=1,2$. The depletion profile on each layer is the difference between the corresponding density profile and the layer bulk density $(1/3)/(2\pi l_B^2)$.
For $q=1$, we observe a full agreement with data from Ref. \cite{Jaworowski_2019} (black dashed lines, labeled JRL2019).}
 \label{Fig:Halperin221:Depletion}
\end{figure}

In Fig.~\ref{Fig:Halperin221:Depletion} we show the depletion profiles for the Halperin state $\Psi_{221}$ in Eq.~\eqref{Eq:Halperin}, which are the ones used to produce Fig~\ref{Fig:Halperin}. Upon adding $q=1$ or $q=2$ quasiholes on layer $A$, the corresponding density decreases (close to the quasihole position), while the density on layer $B$ becomes larger; this is a consequence of the anticorrelations between particles in the two layers, encoded in $\Psi_{221}$. If the quasihole is of the type $\Gamma=AB$, full $A \leftrightarrow B$ symmetry of the density profile is restored -- as we correctly observe in our numerical data.

For a single quasihole ($q=1$) with $\Gamma=A$ or $AB$, we compare our depletion profiles with the ones of Ref.~\cite{Jaworowski_2019}; those are obtained by exact diagonalization of a microscopic model for a bilayer system of particles on the torus, with contact interactions and in the presence of a magnetic field. By adding the appropriate number of flux quanta, and by including a zero-range external potential that pins the quasihole, the authors are able to identify states corresponding to $(\Gamma,q)=(A,1)$ and $(AB,1)$. The corresponding depletion profiles are in perfect agreement with the ones that we obtain for the Halperin state.

\section{Laughlin quasihole and the angular-momentum rigid shift}
\label{App:Laughlin:AFI}

In Section \ref{Sec:AFI} of the main text we showed that the Laughlin quasihole state in Eq.~\eqref{Eq:Laughlin} and the state obtained via angular-momentum rigid shift have the same statistical properties, but different values for the spin $J$. This is a direct consequence of the difference in their depletion profiles, which thus have different quadrupole moments.

\begin{figure}[t]
\includegraphics[width=0.98\linewidth]{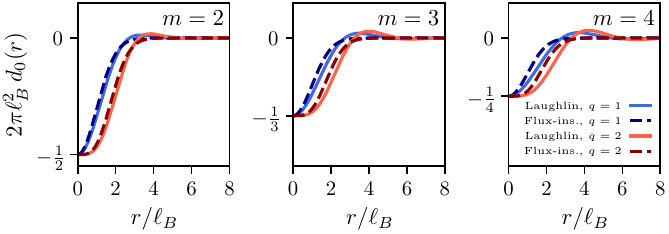}
 \caption{Depletion profiles for the Laughlin state with filling $1/m$, for quasiholes with $q=1,2$, as obtained by sampling $\Psi_{m,q}$ or by Eq.~\eqref{Eq:Laughlin:depletion_with_flux_insertion}.
 }
 \label{Fig:Laughlin:Depletion}
\end{figure}

The depletion profile due to  $q\geq 0$ rigid shifts of the angular-momentum occupation numbers reads
\begin{equation}
d_{0}(z) =
-\bar{\rho}
e^{-\frac{r^2}{2l_B^2}}
\left[ \sum_{l=0}^{q-1}
\frac{1}{l!}
\left( \frac{r^2}{2l_B^2} \right)^l
\right],
\label{Eq:Laughlin:depletion_with_flux_insertion}
\end{equation}
where $\bar{\rho} = 1/(2\pi m l_B^2)$ is the bulk density of the Laughlin state. This expression follows from the combination of the wave functions for different angular momentum states in the LLL, and from the angular-momentum occupation numbers in Eq.~\eqref{Eq:AMSON}.
In Fig.~\ref{Fig:Laughlin:Depletion} we compare this flux-insertion depletion profile with the depletion profile of $\Psi_{m,q}$ obtained by Monte Carlo sampling (that is, the one used to produce Fig.~\ref{Fig:Laughlin}). Both for $q=1$ and $q=2$ we notice a clear difference between the two states, and for instance this rigid-shift depletion profile decays monotonously towards 0, without the damped oscillations which are present for the Laughlin quasihole state.

\section{Fictitious forces and electromagnetic field in a rotating reference frame}
\label{Appendix:RotatingFrame}

The study of a Hall setup in a rotating frame, dubbed $R_2$, plays a crucial role in Sec.~\ref{Sec:AdiabaticRotation}. 
The dynamics in $R_2$ is summarised by Eq.~\eqref{Eq:Ham:R2} and the goal of this appendix is to show that the term $- \frac{\vartheta_f}{T} \hat L_z$ added according to standard prescriptions correctly describes the electromagnetic field in the rotating frame. For simplicity, in this appendix we define $\Omega = \vartheta_f /T$.

Consider first an inertial reference frame with electric and magnetic fields $\mathbf E(\mathbf r)$ and $\mathbf B(\mathbf r)$, and a non-inertial frame that is rotating with constant angular frequency $\mathbf \Omega$. 
In the rotating frame the electric and magnetic fields read (we assume that all relativistic effects can be neglected):
\begin{subequations}
\begin{align}
 \mathbf E' =& \mathbf E+ (\mathbf \Omega \wedge \mathbf r) \wedge \mathbf B ; \\
 \mathbf B' =& \mathbf B- (\mathbf \Omega \wedge \mathbf r) \wedge \mathbf E.
\end{align}
\end{subequations}

Let us specify these formulas to our problem.
In the inertial frame we only have a static and homogeneous magnetic field $\mathbf B_0$; thus, in the rotating frame we have the same homogeneous magnetic field $\mathbf B' = \mathbf B_0$ and an additional radial electric field. In the case that is relevant for the present article, the angular frequency $\mathbf \Omega$ is (i) constant, (ii) parallel to $\mathbf B_0$, and (iii) aligned perpendicularly to the two-dimensional plane where the particles are located; we obtain $\mathbf E' (\mathbf r) = \Omega B_0 r \hat e_{r}$. The scalar potential reads $V'(\mathbf r) = - \Omega \frac{ B_0}{2} r^2$ and the potential energy for a particle with charge $+e$ is
$U'(r) = -  \Omega \frac{r^2}{2 \ell_B^2} $.

Let us now get back to Eq.~\eqref{Eq:Ham:R2}. 
As far as the magnetic field is concerned, $\hat H_{\rm FQHE}$ is the same and thus the magnetic field has not changed.
Let us now consider the term $-  \Omega \hat{L}_z$. We reintroduce for simplicity the notation of classical analytical mechanics
and recall that $\mathbf L$ is the \textit{canonical} angular momentum, to be expressed as $\mathbf L = \mathbf r \wedge (m \mathbf v + e \mathbf A) = \mathbf l + e \mathbf r \wedge \mathbf A$, where $\mathbf l$ is the \textit{mechanical} angular momentum. The term $- \mathbf \Omega \cdot \mathbf l$ is responsible for the mechanical fictious forces that appear in the rotating reference frame. The term $- e \mathbf \Omega \cdot (\mathbf r \wedge \mathbf A)$ is instead responsible for the appearance of the radial electric field. With little algebra:
\begin{equation}
 - e \Omega \left( \mathbf r \wedge \mathbf A \right)_z = - \Omega \frac {eB}{2} r^2 = -  \Omega \frac{r^2}{2 \ell_B^2}.
\end{equation}

\bibliography{LLLquasiholes_v10_arXiv_FINAL.bib}

\end{document}